\documentstyle[aps,floats,eqsecnum,twocolumn,epsf,psfrag,graphicx]{revtex}
\begin{document}
\draft
%-----------------------------------------------------------------------
% for two column style turn on
\twocolumn[\hsize\textwidth\columnwidth\hsize\csname
            @twocolumnfalse\endcsname
%-----------------------------------------------------------------------

\title{Quasi-Spherical Approximation for Rotating Black Holes}

\author{Hisa-aki Shinkai and Sean A. Hayward}
\address{
{\tt shinkai@gravity.phys.psu.edu}, {\tt hayward@gravity.phys.psu.edu}
\\
Centre for Gravitational Physics and Geometry,
104 Davey Laboratory,
The Pennsylvania State University, \\
University Park, PA 16802-6300, U.S.A.
}
\date{October 9, 2000  ~~ revised version ~~ gr-qc/0008075}
%\date{August 30, 2000}

\maketitle

%%%%%%%%%%%%%%%%%%%%%%%%%%%%%%%%%%%%%%%%%%%%%%%%%%%%%%%%%%%%%%%%%%%%%%%
%----------------------------------------------------------------------
% for two column style turn on
\widetext
%----------------------------------------------------------------------
\begin{abstract}
We numerically implement a quasi-spherical approximation scheme
for computing gravitational waveforms for coalescing black holes,
testing it against angular momentum by applying it to Kerr black holes.
As error measures,
we take the conformal strain and specific energy
due to spurious gravitational radiation.
The strain is found to be monotonic rather than wavelike.
The specific energy is found to be at least an order of magnitude smaller than
the 1\% level expected from typical black-hole collisions,
for angular momentum up to at least 70\% of the maximum,
for an initial surface as close as $r=3m$.
\end{abstract}

\pacs{PACS numbers: 04.25.-g, 04.25.Dm, 04.70.Bw, 04.20.Ha, 04.30.-w \\
Preprint numbers: gr-qc/0008075, CGPG-00/8-1}
%----------------------------------------------------------------------
% for two column style turn on
\vskip 2pc]
\narrowtext
%----------------------------------------------------------------------
%%%%%%%%%%%%%%%%%%%%%%%%%%%%%%%%%%%%%%%%%%%%%%%%%%%%%%%%%%%%%%%%%%%%%%%
\section{Introduction}
%%%%%%%%%%%%%%%%%%%%%%%%%%%%%%%%%%%%%%%%%%%%%%%%%%%%%%%%%%%%%%%%%%%%%%%
A quasi-spherical approximation scheme
in a 2+2 decomposition of the space-time
has recently been introduced \cite{qs}.
This proposal is
with the aim of providing a computationally inexpensive estimate
of the gravitational waveforms produced by a black-hole or neutron-star
collision,
given a full numerical simulation up to (or close to) coalescence,
or an analytical model thereof.

The scheme truncates the Einstein equations
by removing second-order terms which would vanish in a spherically
symmetric space-time, cf.\ Bishop {\it et al.}\cite{BGLW}.
Thus when the linearized fields vanish, spherical symmetry is
recovered in full.
Unlike previous work on null-temporal formulations\cite{BGLW,Lehner},
a dual-null formulation is adopted here,
i.e.\ a decomposition of the space-time
by two intersecting foliations of null hypersurfaces.
Technical advantages of the scheme include that
only ordinary differential equations need be solved,
and that the dual-null formulation is adapted to radiation extraction.
Advantages of applicability include that
no prescribed background is required
and that arbitrarily rapid dynamical processes
(close to spherical symmetry) are allowed.
The pressing question concerns
how well the scheme handles deviations from spherical symmetry.
The principal such deviation in the context of coalescing black holes
is expected to be due to angular momentum.
The primary test case is therefore Kerr black holes,
the unique stationary vacuum black holes.

This article reports a numerical implementation
of the quasi-spherical approximation and its application to Kerr black holes,
taking Boyer-Lindquist quasi-spheres.
Except in the non-rotating (Schwarzschild) case,
the approximation produces spurious gravitational radiation,
not present in the exact solution.
We consider two measures of the error introduced by the approximation.
Firstly, the practical measure is the waveform of the spurious strain,
as compared to signals expected to be measured by interferometers.
Secondly, we measure the specific energy,
i.e.\ the ratio of the radiated energy $E$ to the original mass $m$;
this is a conservative measure,
as it involves summing the errors over all angles of the sphere,
whereas observation is restricted to a particular angle.

In principle these quantities depend on only the spin parameter $a/m$,
the relative initial radius $r_0/m$ and, in the case of the strain, the angle.
For the approximation to be useful, the error should be significantly less than
the values expected for a realistic black-hole collision.
Typical values for the specific energy obtained from numerical simulations
\cite{headonNR} or
from the close-limit approximation \cite{closelimit}
have increased from early estimates to around 1\%
if the initial relative momentum\cite{baker}
or angular momentum\cite{closelimit2} is appreciable.
The theoretical limit on how badly an approximation might perform
is much higher:
29\% of the mass of a maximally rotating Kerr black hole
may be extracted by the Penrose process \cite{wald}.

The article is organized as follows. Section  \ref{sec_formapprox}
describes the dual-null formalism, the quasi-spherical approximation
and the observables, strain and energy.
Section \ref{sec_modelnumeric} describes our model and numerical
integration procedures.
The numerical results are shown in Section \ref{sec_result}
and we summarize the article in  Section \ref{sec_conclude}.

%%%%%%%%%%%%%%%%%%%%%%%%%%%%%%%%%%%%%%%%%%%%%%%%%%%%%%%%%%%%%%%%%%%%%%%
\section{Formulation and Approximation}\label{sec_formapprox}
%%%%%%%%%%%%%%%%%%%%%%%%%%%%%%%%%%%%%%%%%%%%%%%%%%%%%%%%%%%%%%%%%%%%%%%
\subsection{Dual-null formulation}
The quasi-spherical approximation\cite{qs}
is based on a dual-null formulation\cite{dn}
of Einstein gravity\cite{dne}, summarized as follows.
One takes two intersecting families of null hypersurfaces labelled by $x^\pm$.
Then the normal 1-forms $n^\pm=-dx^\pm$ satisfy
\begin{equation}
g^{-1}(n^\pm,n^\pm)=0
\end{equation}
where $g$ is the space-time metric.
The relative normalization of the null normals
may be encoded in a function $f$ defined by
\begin{equation}
e^f=-g^{-1}(n^+,n^-).
\end{equation}
Then the induced metric on the transverse surfaces,
the spatial surfaces of intersection, is found to be
\begin{equation}
h=g+2e^{-f}n^+\otimes n^-,
\end{equation}
where $\otimes$ denotes the symmetric tensor product.
The covariant derivative of $h$ is denoted by $D$.
The dynamics is described by Lie transport
along two commuting evolution vectors $u_\pm$:
\begin{equation}
[u_+,u_-]=0.
\end{equation}
Specifically, the evolution derivatives,
to be discretized in a numerical code, are
\begin{equation}
\Delta_\pm=\bot L_{u_{\pm}},
\end{equation}
where $\bot$ indicates projection by $h$ and $L$ denotes the Lie derivative.
There are two shift vectors
\begin{equation}
s_\pm=\bot u_\pm.
\end{equation}
In a coordinate basis $(u_+,u_-;e_i)$ such that
$u_\pm=\partial/\partial x^\pm$,
where $e_i=\partial/\partial x^i$ is a basis for the transverse surfaces,
the metric takes the form
\begin{eqnarray}
g&=&h_{ij}(dx^i+s_+^idx^++s_-^idx^-)\otimes\nonumber\\
&&(dx^j+s_+^jdx^++s_-^jdx^-)
-2e^{-f}dx^+\otimes dx^-.
\end{eqnarray}
Then $(h,f,s_\pm)$ are configuration fields
and the independent momentum fields are found to be linear combinations of
\begin{eqnarray}
\theta_\pm&=&*L_\pm{*}1, \\
\sigma_\pm&=&\bot L_\pm h-\theta_\pm h, \\
\nu_\pm&=&L_\pm f, \\
\omega&=&\textstyle{1\over2}e^fh([l_-,l_+]),
\end{eqnarray}
where $*$ is the Hodge operator of $h$
and $L_\pm$ is shorthand for the Lie derivative along the null normal vectors
\begin{equation}
l_\pm=u_\pm-s_\pm=e^{-f}g^{-1}(n^\mp).
\end{equation}
Then the functions $\theta_\pm$ are the expansions,
the traceless bilinear forms $\sigma_\pm$ are the shears,
the 1-form $\omega$ is the twist,
measuring the lack of integrability of the normal space,
and the functions $\nu_\pm$ are the inaffinities,
measuring the failure of the null normals to be affine.
The fields $(\theta_\pm,\sigma_\pm,\nu_\pm,\omega)$
encode the extrinsic curvature of the dual-null foliation.
These extrinsic fields are unique up to duality $\pm\mapsto\mp$
and diffeomorphisms which relabel the null hypersurfaces, i.e.\
$dx^\pm\mapsto e^{\lambda_\pm}dx^\pm$
for functions $\lambda_\pm(x^\pm)$.

It is also useful to decompose $h$ into
a conformal factor $\Omega$ and a conformal metric $k$ by
\begin{equation}
h=\Omega^{-2}k,
\end{equation}
such that
\begin{equation}
\Delta_\pm\hat{*}1=0,
\end{equation}
where $\hat{*}$ is the Hodge operator of $k$,
satisfying ${*}1=\hat{*}\Omega^{-2}$.
Taking quasi-spherical coordinates $x^i=(\theta,\phi)$ such that
$\hat{*}1=\sin\theta d\theta\wedge d\phi$,
$\Omega^{-1}$ is the quasi-spherical radius.
In an asymptotically flat space-time,
it becomes convenient to use the conformally rescaled expansions and shears
\begin{eqnarray}
\vartheta_\pm&=&\Omega^{-1}\theta_\pm, \\
\varsigma_\pm&=&\Omega\sigma_\pm,
\end{eqnarray}
since they are finite and generally non-zero at null infinity $\Im^\mp$.

\subsection{Quasi-spherical approximation}
Of the dynamical fields and operators introduced above,
$(s_\pm,\sigma_\pm,\omega,D)$ vanish in spherical symmetry,
while $(h,f,\theta_\pm,\nu_\pm,\Delta_\pm)$ generally do not.
The quasi-spherical approximation consists of linearizing in
$(s_\pm,\sigma_\pm,\omega,D)$,
i.e.\ setting to zero any second-order terms in these quantities.
This yields a greatly simplified truncation \cite{qs}
of the full field equations,
the first-order dual-null form of the vacuum Einstein system\cite{dne}.
In particular, the truncated equations decouple into a three-level hierarchy,
the last level being irrelevant to determining the gravitational waveforms.
The remaining equations are the quasi-spherical equations
\begin{eqnarray}
\Delta_\pm\Omega&=&-\textstyle{1\over2}\Omega^2\vartheta_\pm, \\
\Delta_\pm f&=&\nu_\pm, \\
\Delta_\pm\vartheta_\pm&=&-\nu_\pm\vartheta_\pm, \\
\Delta_\pm\vartheta_\mp
&=&-\Omega(\textstyle{1\over2}\vartheta_+\vartheta_-+e^{-f}), \\
\Delta_\pm\nu_\mp
&=&-\Omega^2(\textstyle{1\over2}\vartheta_+\vartheta_-+e^{-f}),
\end{eqnarray}
and the linearized equations
\begin{eqnarray}
\Delta_\pm k&=&\Omega\varsigma_\pm, \\
%\Delta_\pm\varsigma_\mp
%&=&-\textstyle{1\over2}\Omega\vartheta_\mp\varsigma_\pm.
%\\
\Delta_\pm\varsigma_\mp
&=&\Omega(\varsigma_+\cdot k^{-1}\cdot\varsigma_-
-\textstyle{1\over2}\vartheta_\mp\varsigma_\pm).\label{eql}
\end{eqnarray}
These are all ordinary differential equations;
no transverse $D$ derivatives occur.
Thus we have an effectively two-dimensional system
to be integrated independently at each angle of the sphere.

The initial-data formulation is based on
a spatial surface $S$ orthogonal to $l_\pm$
and the null hypersurfaces $\Sigma_\pm$ generated from $S$ by $l_\pm$,
assumed future-pointing.
The initial data for the above equations are
$(\Omega,f,k,\vartheta_\pm)$ on $S$
and $(\varsigma_\pm,\nu_\pm)$ on $\Sigma_\pm$.
We will take $l_+$ and $l_-$ to be outgoing and ingoing respectively.

\subsection{Strain}
The variables are directly related to physically measurable quantities.
In particular,
\begin{equation}
\epsilon={1\over2}\int_\gamma\dot kd\tau
\end{equation}
is the transverse strain tensor measured along a worldline $\gamma$
normal to the transverse surfaces,
where $\dot k=\bot L_\lambda k$
in terms of a vector $\lambda=\partial/\partial\tau$ tangent to $\gamma$.
For a detector at large distance,
one may apply the linearized approximation,
where $2\epsilon$ reduces to the transverse traceless metric perturbation
of a linearized plane gravitational wave.
In a weak gravitational field, one may use Newtonian physics,
where $\epsilon$ reduces to the Newtonian strain tensor.
Thus the displacements to be measured by an interferometer are
\begin{equation}
{\delta\ell\over\ell}=\epsilon(e,e)
\end{equation}
where the unit vector $e$ is the direction of displacement.

Writing $\lambda=a^+l_++a^-l_-$ yields
\begin{equation}
\epsilon={1\over2}\int_\gamma\Omega(a^+\varsigma_++a^-\varsigma_-)d\tau.
\end{equation}
Since the strain vanishes at future null infinity $\Im^+$,
it is convenient to use the conformal strain tensor
\begin{equation}
\varepsilon={1\over2}\int \varsigma_-dx^-,
\label{conformalstrain}
\end{equation}
where the integal is at constant $x^+$.
We will denote its plus and cross components by
$\varepsilon_+ = \varepsilon_{\theta\theta}$ and
$\varepsilon_\times = \varepsilon_{\theta\phi}$.
In order to compare with observational results,
one converts back to the strain,
\begin{equation}
\epsilon={\varepsilon\over{R}},
\label{scale_of_strain}
\end{equation}
where $R$ is the distance between the source and the detector.

\subsection{Energy}
We define the energy flux $\phi$ of the gravitational waves,
or more conveniently, the conformal energy flux $\varphi=\Omega^{-2}\phi$,
as the 1-form $\varphi=\varphi_+dx^++\varphi_-dx^-$, where
\begin{equation}
\varphi_\pm=-{e^f\vartheta_\mp k^{ab}k^{cd}\varsigma_{\pm ac}\varsigma_{\pm bd}
\over{64\pi}}.
\label{flux}
\end{equation}
These expressions have the same form as those for
the conformal Bondi flux at $\Im^\mp$\cite{qs},
but we propose using them locally.
Then $\phi_-$ is the outgoing flux
and $\phi_+$ is the ingoing flux.
The corresponding energy $E$ of the gravitational waves is then given by
\begin{equation}
\Delta_\pm E=  \oint\hat{*}\varphi_\pm
\label{energy}
\end{equation}
with the initial condition $E|_S=0$.
Thus the Bondi energy at $\Im^+$ is $E+E_0$,
where $E_0$ is the Bondi energy at the intersection with $\Sigma_+$,
which in the Kerr case will be just the ADM mass $m$.
We propose using the specific energy $E/m$ at $\Im^+$
as a measure of the strength of a black-hole collision,
following various references\cite{headonNR,closelimit,baker,closelimit2}.
This is the fraction of the original mass-energy which has radiated away.
%%%%%%%%%%%%%%%%%%%%%%%%%%%%%%%%%%%%%%%%%%%%%%%%%%%%%%%%%%%%%%%%%%%%%%%
\section{Model and Numerical Procedures}\label{sec_modelnumeric}
%%%%%%%%%%%%%%%%%%%%%%%%%%%%%%%%%%%%%%%%%%%%%%%%%%%%%%%%%%%%%%%%%%%%%%%
\subsection{Model: Kerr black hole}
\begin{figure}[tbh]
   \unitlength=1in
          \vspace*{0.1in}
% \centerline{\epsfxsize=5cm \epsfbox{qs1.eps}}
%  \begin{center}
        \psfragscanon
         \psfrag{SigM}{$\Sigma_-$}
         \psfrag{SigP}{$\Sigma_+$}
         \psfrag{rrbegin}{$r=\lambda r_H$}
         \psfrag{rplus1}{$r=r_H$}
         \psfrag{xm}{$ x^-$}
         \psfrag{xp}{$ x^+$}
         \psfrag{rrend}{$r=n r_H$}
         \psfrag{r0}{$S: r=r_0$}
         \psfrag{scriplus}{$\Im^+$}
         \psfrag{scriminus}{$\Im^-$}
         \psfrag{infplus}{$i^+$}
         \psfrag{infzero}{$i^0$}
             \makebox[\columnwidth]{
              \includegraphics[height=3.0in,keepaspectratio,clip]{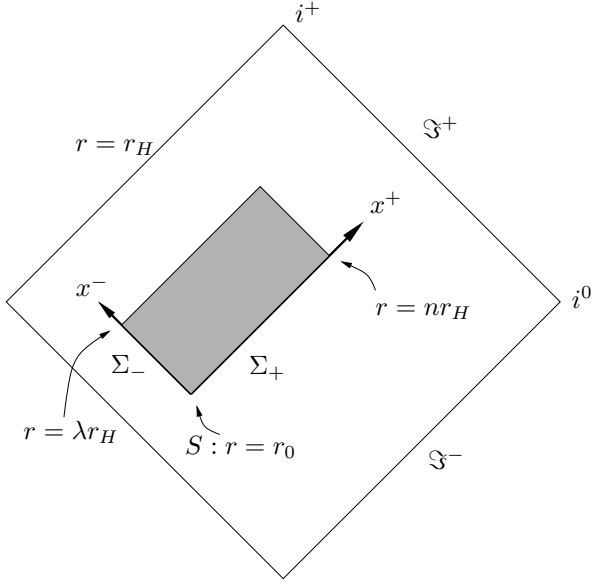}
%          \hspace*{0.2in}
                }
%    \end{center}
          \vspace*{0.1in}
\caption{The region of numerical integration is shown as the shaded region
in the picture.
Initial data is prescribed on a spatial surface $S$
of constant Boyer-Lindquist $r=r_0$ and $t$,
and the null hypersurfaces $\Sigma_\pm$
generated from it.
On $\Sigma_-$ ($\Sigma_+$), the $x^-$ ($x^+$)
coordinate is set so as to cover
the region $\lambda r_H \le r \le r_0$ ($r_0 \le r \le n r_H$),
where $1<\lambda\approx1$ and $n \gg 1 $ are
constants to be set by hand.
}
\label{fig1}
\end{figure}

As our model, we take a Kerr black-hole geometry,
\begin{eqnarray}
ds^2&=&-{\Delta \over \Sigma}[dt-a \sin^2\theta d\phi]^2
+{\sin^2\theta\over\Sigma}[(r^2+a^2)d\phi-adt]^2
\nonumber \\&&
+{\Sigma\over\Delta}dr^2+\Sigma d\theta^2,
\label{KerrBL}
%\\
%&=&-{\Delta-a^2\sin^2\theta \over \Sigma}dt^2 + {2a\sin^2\theta\over\Sigma}
%[\Delta-(r^2+a^2)]dtd\phi+{\Sigma\over\Delta}dr^2 \nonumber \\
%&~&+\Sigma d\theta^2+{\sin^2\theta\over\Sigma}[(r^2+a^2)^2
%-a^2\Delta\sin^2\theta]d\phi^2 \\
%&=& -(1+{q^2 -2mr \over \Sigma}) dt^2
%+{2a \sin^2\theta \over \Sigma}(q^2-2mr) dt d\phi
%+{\Sigma \over \Delta} dr^2  \nonumber \\
%&&+\Sigma d\theta^2 + R^2 \sin^2\theta d\phi^2
\end{eqnarray}
where
\begin{eqnarray}
\Delta&=&r^2-2mr+a^2 \\
\Sigma&=&r^2+a^2\cos^2\theta
\end{eqnarray}
and $m$ is the mass and $am$ the angular momentum.
The horizon radius is denoted by $r_H=m+\sqrt{m^2-a^2}$.
We take the quasi-spherical approximation
adapted to these (Boyer-Lindquist) coordinates,
i.e.\ the initial surface $S$ is of constant $r=r_0$ and constant $t$,
as depicted in Fig.\ref{fig1}.
The inaccuracy of the quasi-spherical approximation as measured by $E/m$
then depends in principle only on $a/m$ and $r_0/m$.
We expect $E/m$ to be monotonically increasing in $a/m$, from zero at $a=0$
to a maximum at $a=m$, since angular momentum is the cause of the asphericity.
Similarly, $E/m$ is expected to be monotonically decreasing in $r_0/m$,
to zero
at infinity, since the approximation should be better at large distances.

%%%%%%%%%%%%%%%%%%%%%%%%%%%%%%%%%% figure %%%%%%%%%%%%%%%%
%%%%%%%%%%%%%%%%%%%%%%%%%%%%%%%%%% figure %%%%%%%%%%%%%%%%

\subsection{Initial data}
An explicit dual-null form of the Kerr metric is not known
except on the symmetry axis \cite{Carter66} or
in the Schwarzschild case, as the Kruskal form.
Although there is an effort for this direction \cite{PI}, we
did not find explicit double-null coordinates which are
well behaved at the outer horizons and infinity,
despite trying changes in angular coordinate and
Kruskal-type rescalings \cite{shinji}.
However, we can construct the initial data analytically
as functions of $(r,\theta)$,
then convert to the required functions of $(x^\pm,\theta)$, as follows.
We remark that our initial surfaces are $\Sigma_\pm$, and the
following method applies outside the horizons.
%However, we can find the initial data analytically
%as functions of $(r,\theta)$,
%then convert to the required functions of $(x^\pm,\theta)$, as follows.

The null normal vectors are initially given by
\begin{equation}
\left.l_\pm\right\vert_{\Sigma_\pm}
=\left({\Sigma\over{2(\Delta-a^2\sin^2\theta)}}\right)^{1/2}\partial_t
\pm\left({\Delta\over{2\Sigma}}\right)^{1/2}\partial_r
\end{equation}
where the normalization is such that
\begin{equation}
\left.f\right\vert_S=0.
\end{equation}
The apparent degeneracy at $\Delta=a^2\sin^2\theta$
is just the boundary of the ergoregion where $\partial_t$ becomes spatial;
the dual-null coordinates extend through.
We also fix
\begin{equation}
\left.\nu_\pm\right\vert_{\Sigma_\pm}=0,
\end{equation}
which means that $x^\pm\vert_{\Sigma_\pm}$ are affine parameters.
This implies $f\vert_{\Sigma_\pm}=0$,
fixing $l_\mp\vert_{\Sigma_\pm}$
and therefore locally determining the dual-null foliation.

The quasi-spherical conformal factor is
\begin{equation}
\Omega=\left((r^2+a^2)^2-\Delta a^2\sin^2\theta\right)^{-1/4}
\end{equation}
which is real and positive.
Then we obtain the conformal metric
\begin{equation}
k = \Omega^2\Sigma d\theta^2
+{\sin^2\theta\over{\Omega^2\Sigma}}d\phi^2,
\\
\end{equation}
and the conformally rescaled expansions and shears
\begin{eqnarray}
\left.\vartheta_\pm\right\vert_S
%&=&\pm\left({\Delta\over{2\Sigma}}\right)^{1/2}\Omega^3U,
&=&\pm\sqrt{\Delta\over{2\Sigma}}\Omega^3U,
\\
\left.\varsigma_\pm\right\vert_{\Sigma_\pm}
%&=&\pm\left({\Delta\over{2\Sigma}}\right)^{1/2}\Omega^5Va^2\sin^2\theta
&=&\pm\sqrt{\Delta\over{2\Sigma}}\Omega^5Va^2\sin^2\theta
\left(d\theta^2-{\sin^2\theta\over{\Omega^4\Sigma^2}}d\phi^2\right),
\end{eqnarray}
%The conformally rescaled expansions and shears are
%$$\vartheta_\pm=-2\Omega^{-2}L_\pm\Omega\qquad
%\varsigma_\pm=\Omega^{-1}\bot L_\pm k.$$
where
\begin{eqnarray}
U&=&-2\Omega^{-5}\partial_r\Omega=2r(r^2+a^2)-(r-m)a^2\sin^2\theta,
\\
V&=&{\Omega^{-6}\partial_r(\Omega^2\Sigma)\over{a^2\sin^2\theta}}
=r^3+3mr^2+a^2(r-m)\cos^2\theta.
\end{eqnarray}

To complete the initial data construction,
we need to know the initial data on $\Sigma_\pm$
as functions of $(x^\pm,\theta)$.
So we need to know $r\vert_{\Sigma_\pm}$ as functions of $(x^\pm,\theta)$.
This is determined by the equations
\begin{equation}
\left.\Delta_\pm r\right\vert_{\Sigma_\pm}
=\pm\left({\Delta\over{2\Sigma}}\right)^{1/2},
\end{equation}
giving
\begin{equation}
\left.x_\pm\right\vert_{\Sigma_\pm}
=\pm\int_{r_0}^r\left({2\Sigma\over{\Delta}}\right)^{1/2}dr', \label{xpm_by_r}
\end{equation}
%gives $\left.x_\pm\right\vert_{\Sigma_\pm}$ as functions of $(r,\theta)$,
where the integral is along a curve of constant $(\theta,\phi)$.
Note that the $\Delta$ factor means that
we take $S$ outside the horizons $r=r_H$,
which anyway is the region of interest.
We numerically integrated (\ref{xpm_by_r})
using a fourth order Runge-Kutta method
(Fehlberg method), then inverted.
This was checked against the analytic solution
in the equatorial plane $\theta=\pi/2$:
\begin{equation}
x^\pm\vert_{\Sigma_\pm}
= \pm\sqrt{2} \left( \sqrt{\Delta} + m \ln (r-m+\sqrt{\Delta}) \right) \mp c.
\label{xpm_from_r}
\end{equation}
As $m$ is an overall scale, we fixed it to unity.

%%%%%%%%%%%%%%%%%%%%%%%%%%%%%%%%%%%%%%%%%%%%%%%%%%%%%%%%%%%%%%%%%%%%%%%
\subsection{Evolution procedures}
Here we describe our numerical procedures.
We have a set of ordinary differential equations in two variables.
As pointed out by
Gundlach and Pullin \cite{GP97}, free evolution schemes
in such a system may lead to unstable evolution.
This fact was also seen in our experience, and we developed a kind of
predictor-corrector scheme similar to that of Hamad\'e and Stewart
\cite{HS96}.

%%%%%%%%%%%%%%%%%%%%%%%%%%%%%%%%%% figure %%%%%%%%%%%%%%%%
\begin{figure}[tbh]
   \unitlength=1in
          \vspace*{0.1in}
% \centerline{\epsfxsize=5cm \epsfbox{qs1.eps}}
%  \begin{center}
        \psfragscanon
         \psfrag{Sigmam}{$\Sigma_-$}
         \psfrag{Sigmap}{$\Sigma_+$}
         \psfrag{ukn}{$u^n_k$}
         \psfrag{uknn}{$u^{n+1}_k$}
         \psfrag{ukkn}{$u^{n+1}_{k-1}$}
         \psfrag{dm}{$\Delta x^-$}
         \psfrag{dp}{$\Delta x^+$}
             \makebox[\columnwidth]{
              \includegraphics[height=1.4in,keepaspectratio,clip]{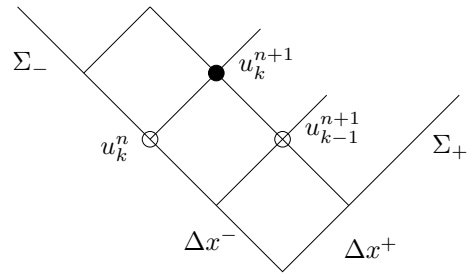}
          \hspace*{0.1in}
                }
%    \end{center}
          \vspace*{0.1in}
\caption{The dual-null integration scheme.
In order to obtain the data at grid $u^{n+1}_k$, we need both
$u^n_k$ and $u^{n+1}_{k-1}$.  }
\label{fig_grid}
\end{figure}
%%%%%%%%%%%%%%%%%%%%%%%%%%%%%%%%%% figure %%%%%%%%%%%%%%%%

The actual steps we took are the followings.
The set of variables is
$u\equiv(\Omega, f, \vartheta_\pm, \nu_\pm,
k_{ab}, \varsigma_\pm{}_{ab})$.
Let us schematically express a set $u$
at a point $x^-=k$ on a slice $\Sigma_- (x^+=n)$
as  $u^n_k$.
The data $u^{n+1}_k$ is determined from both $u^{n+1}_{k-1}$ and
$u^{n}_{k}$ as in Fig. \ref{fig_grid}.
Suppose we have already all the data at $u^{n+1}_{k-1}$ and
$u^{n}_{k}$.

\begin{itemize}
\item[(1)] Firstly, we evolve along the $x^+$-direction, say from
$u^{n}_{k}$ to $u^{n+1}_k$.  We have a set of equations for
$(\Omega, f, \vartheta_\pm, \nu_-, k_{ab}, \varsigma_-{}_{ab})$,
\begin{eqnarray}
\Delta_+ \Omega &=&
%\p^k \ptl_k \Omega
-\textstyle{1\over2}\Omega^2\vartheta_+,
\\
\Delta_+ f
  &=&  %\p^k \partial_k f +
\nu_+,
\\
\Delta_+ \vartheta_+ &=&  %\p^k \ptl_k \vartheta_+
-\nu_+\vartheta_+,   \\
\Delta_+ \vartheta_- &=&  %\p^k \ptl_k \vartheta_-
-\Omega(\textstyle{1\over2}\vartheta_+\vartheta_-+e^{-f}),  \\
\Delta_+ \nu_- &=&  %\p^k \ptl_k \nu_-
-\Omega^2(\textstyle{1\over2}\vartheta_+\vartheta_-+e^{-f}), \\
%h_i^a h_j^b\Delta
\Delta_+ k_{ab} &=&
%h_i^a h_j^b ( \p^k \ptl_k k_{ab} - 2 k_{k(a} \ptl_{b)} \p^k)+
\Omega\varsigma_+{}_{ab}, \\
%%h_i^a h_j^b
%\Delta_+ \varsigma_-{}_{ab}&=&
%% h_i^a h_j^b( \p^k \ptl_k \varsigma_-{}_{ab}
%% - 2 \varsigma_-{}_{k(a} \ptl_{b)} \p^k)
%-\textstyle{1\over2}\Omega\vartheta_-\varsigma_+{}_{ab}.
%\\
\Delta_+\varsigma_-{}_{ab}
&=&\Omega(\varsigma_+{}_{ac}k^{cd}\varsigma_-{}_{db}
-\textstyle{1\over2}\vartheta_-\varsigma_+{}_{ab}).\label{eql}
\end{eqnarray}
The step is integrated using the Fehlberg method.
Note that  we do not have equations for evolving
  $\nu_+$ and $\varsigma_+$, therefore we have to interpolate them
using
$(\nu_+, \varsigma_+{}_{ab})^n_{k}$ and
$(\nu_+, \varsigma_+{}_{ab})^{n+1}_{k}$.
The latter was linearly extrapolated for the first iteration,
but will be updated after an integration along the $x^-$-direction
(next step) has been done.

\item[(2)] Secondly, we evolve along the $x^-$-direction, from
$u^{n+1}_{k-1}$ to $u^{n+1}_k$.  We have a set of equations for
$( \nu_+, \varsigma_+{}_{ab})$,
\begin{eqnarray}
\Delta_- \nu_+ &=& %  \m^k \ptl_k \nu_+
-\Omega^2(\textstyle{1\over2}\vartheta_+\vartheta_-+e^{-f}),  \\
%%h_i^a h_j^b
%\Delta_- \varsigma_+{}_{ab} &=&
%%h_i^a h_j^b ( \m^k \ptl_k \varsigma_+{}_{ab}
%%- 2 \varsigma_+{}_{k(a} \ptl_{b)} \m^k)
%-\textstyle{1\over2}\Omega\vartheta_+\varsigma_- {}_{ab},
%\\
\Delta_-\varsigma_+{}_{ab}
&=&\Omega(\varsigma_+{}_{ac}k^{cd}\varsigma_-{}_{db}
-\textstyle{1\over2}\vartheta_+\varsigma_-{}_{ab}),
\label{eql}
\end{eqnarray}
for completing the set $u$, but we also evolve
$\vartheta_\pm, \Omega$ and $f$ by
\begin{eqnarray}
\Delta_- \Omega &=&
%\m^k \ptl_k \Omega
-\textstyle{1\over2}\Omega^2\vartheta_-
\\
\Delta_- f
  &=& % \m^k \ptl_k f +
\nu_-,
\\
\Delta_- \vartheta_- &=&  %\m^k \ptl_k \vartheta_-
-\nu_-\vartheta_-,  \\
\Delta_- \vartheta_+ &=&  %\m^k \ptl_k \vartheta_+
-\Omega(\textstyle{1\over2}\vartheta_+\vartheta_-+e^{-f}),
\end{eqnarray}
Here again
we have to interpolate $\varsigma_-$ and $\nu_-$
in integrating the above, and we use a cubic spline
interpolation
using
$(\nu_-, \varsigma_-{}_{ab})^{n+1}_{k_i}$ ($1\leq k_i\leq k$),
where the data $(\nu_-, \varsigma_-{}_{ab})^{n+1}_{k}$
was given in the previous step (1).

\item[(3)] We check the
consistencies of the evolution, by monitoring the differences of
$(k_{ab}, \vartheta_\pm, \Omega, f)^n_k$ from the above step (1) and (2).
If they are all within a tolerance, then we finish this evolution step
by updating
$(\nu_+, \varsigma_+{}_{ab}, \vartheta_\pm, \Omega, f)$ as a value
at $u^{n+1}_{k}$.  If not, we repeat back to the step (1).
\end{itemize}

We construct a numerical grid in $x^\pm$ space
with constant spacing in each direction.
The iteration procedures are completed a couple of times at each grid point.
The results shown in this article are obtained
by setting the tolerance in the above step (3) to $10^{-5}$.
The code was tested for the Schwarzschild case,
for which the analytic expression
in dual-null coordinates is known;
the calculated expansion $\vartheta_\pm$
differed from the exact expression to within $10^{-6}$.

In the next section, we present our evolutions of a Kerr black-hole
space-time under this quasi-spherical approximation.  We chose
the initial null slice $\Sigma_-$ so as to
cover the region
$1.25 r_H < r < r_0$.
We stopped
the evolution at $x^+=30$, which corresponds to $r$ being $25 \sim 30m$,
depending on the values of $r_0$ and $a$.
We took 51 grid points in the $x^-$ direction
and 11 grid points in $\theta=[0, \pi/2]$,
and evolved with grid
separation $\Delta x^+ = 0.5 \Delta x^-$.

%%%%%%%%%%%%%%%%%%%%%%%%%%%%%%%%%%%%%%%%%%%%%%%%%%%%%%%%%%%%%%%%%%%%%%%
\section{Numerical Results}\label{sec_result}
%%%%%%%%%%%%%%%%%%%%%%%%%%%%%%%%%%%%%%%%%%%%%%%%%%%%%%%%%%%%%%%%%%%%%%%

%%%%%%%%%%%%%%%%%%%%%%%%%%%%%%%%%% figure %%%%%%%%%%%%%%%%
\begin{figure}[t]
%\begin{figure}[tbh]
\setlength{\unitlength}{1in}
%  \centerline{\epsfxsize=5cm \epsfbox{logEsum.eps}}\hspace*{0.4in}
\begin{picture}(3.30,5.2)
\put(0.10,2.9){\epsfxsize=3.0in \epsfysize=1.78in \epsffile{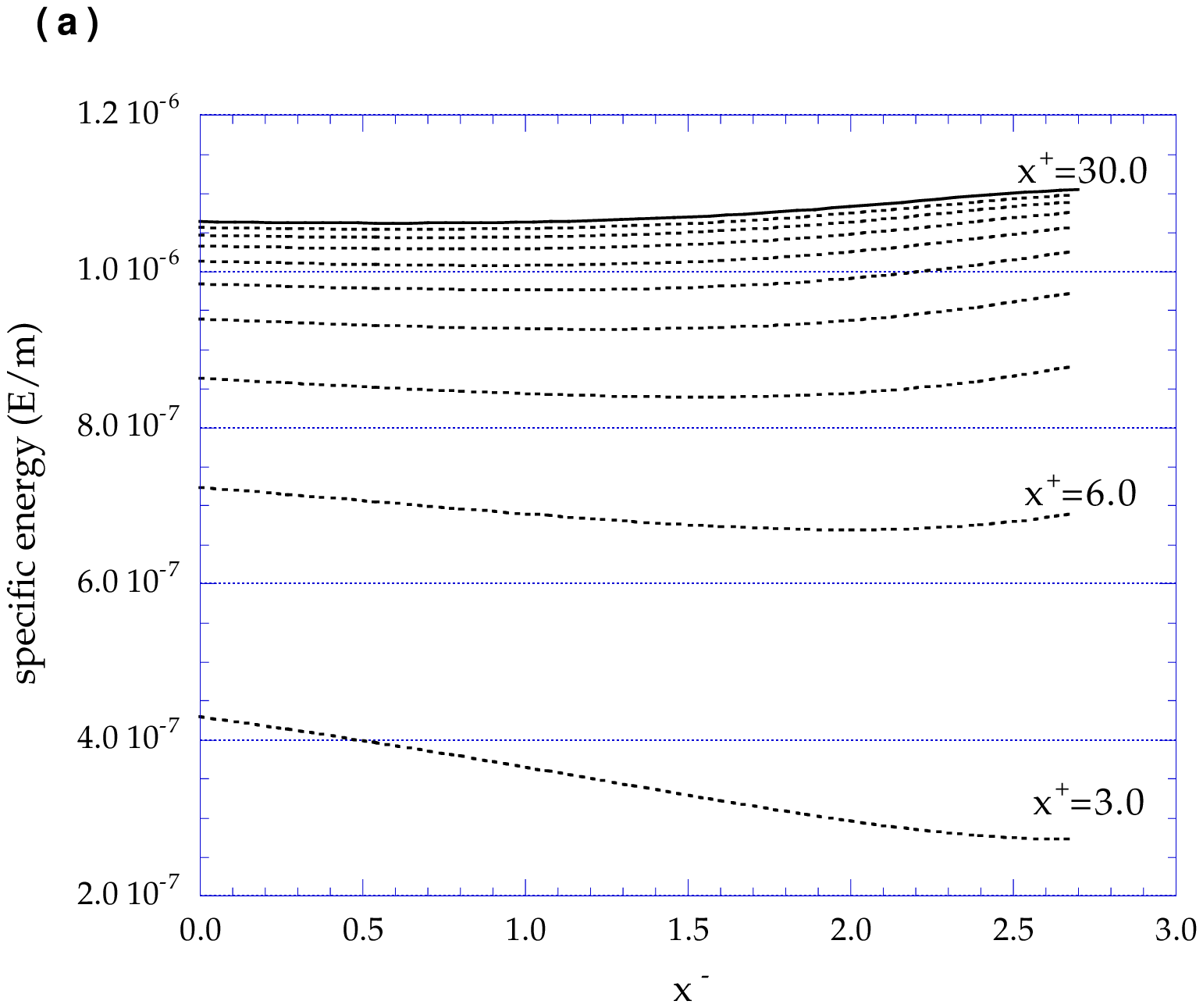} }
\put(0.10,0.2){\epsfxsize=3.0in \epsfysize=1.78in \epsffile{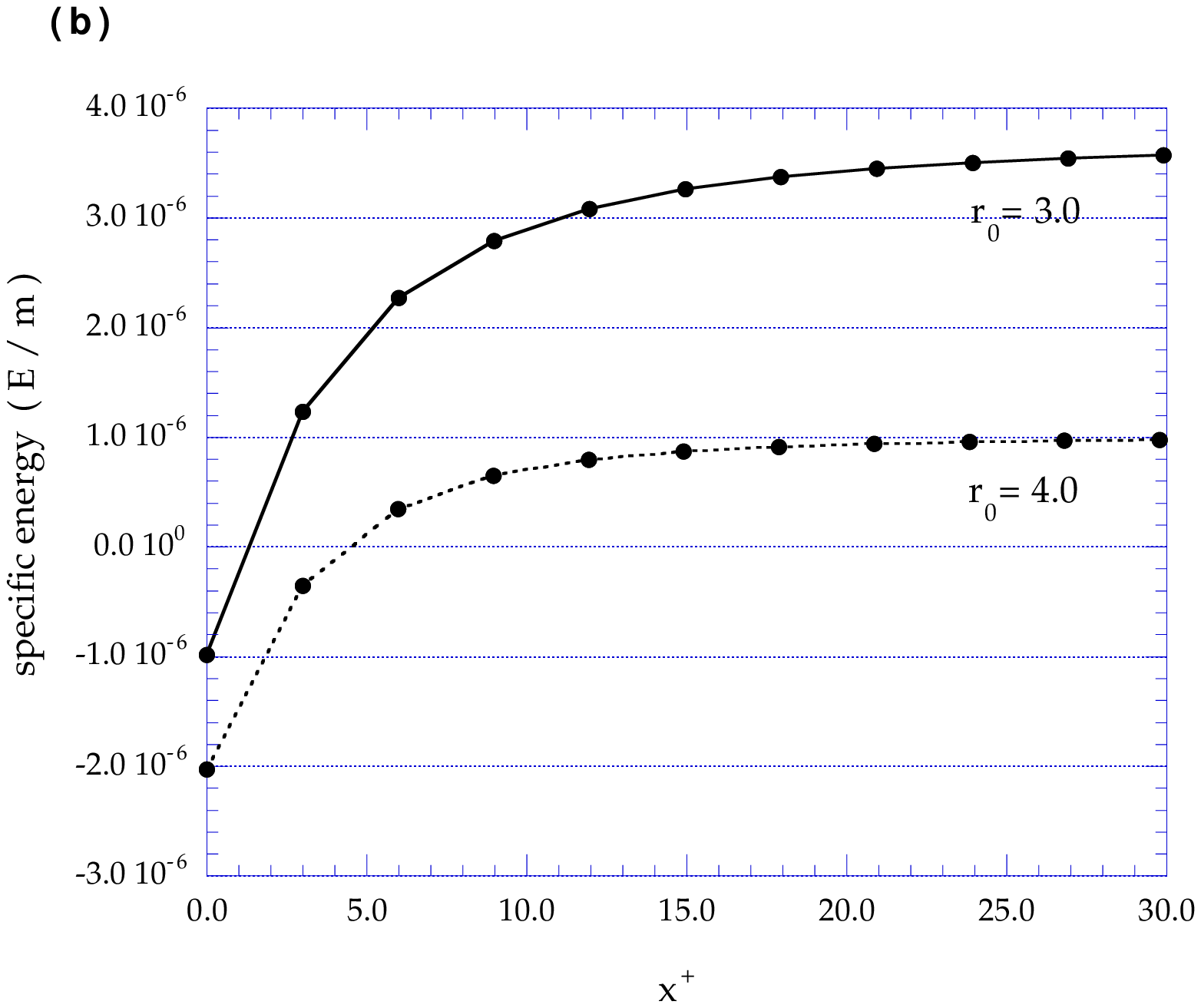} }
\end{picture}

\caption{Specific energy $E/m$ for $a/m=0.1$.
(a) $E/m$ is plotted as a function of the ingoing null coordinate $x^-$
for each constant outgoing null coordinate $x^+=0.0, 3.0, \cdots, 30.0$.
We set $r_0=4.0$ for this plot.
(b) The integrated $E/m$ over the ingoing null coordinate $x^-$
is shown as a function of the outgoing null coordinate $x^+$.
We show both $r_0=3.0$ and 4.0 cases.
We see that the specific energy
converges to a particular positive value in the $x^+$ direction,
as expected.}
\label{fig_logE}
\end{figure}
%%%%%%%%%%%%%%%%%%%%%%%%%%%%%%%%%% figure %%%%%%%%%%%%%%%%

Recall that our principal measures of the gravitational radiation are
the conformal strain $\varepsilon$ (\ref{conformalstrain})
and the specific energy $E/m$ (\ref{energy}).
Since we are using conformal variables, we expect that we can evolve
towards the asymptotically flat region without long-term evolution
in the $x^+$ direction.
In Fig.\ref{fig_logE}, we plotted the specific energy $E/m$
at the boundaries of the integration region.
We integrated the $\Delta_+ E$ equation of (\ref{energy}) along the
hypersurface $\Sigma_+$ ($x^-=0$), setting $E=0$ on $S$ ($x^+=x^-=0$),
then integrated $E$ using the $\Delta_- E$ equation
at each constant $x^+$.
We plotted $E$ as a function of
$x^-$ at a constant $x^+$ surface in Fig.\ref{fig_logE}(a).
We see that $E$ is converging to a line (the solid line in the figure),
and not diverging even close  to the black hole (at larger $x^-$).
Fig.\ref{fig_logE}(b) plots $E$ at the final value of $x^-$
as a function of $x^+$.
%The plot is for
%Kerr spacetime with $a=0.1$, and $r_0=3.0$ and 4.0.
We see from the figure that the energy measured
for increasing $x^+$ converges at some value,
as expected.

\begin{figure}[t]
\setlength{\unitlength}{1in}
%  \centerline{\epsfxsize=5cm \epsfbox{strain+.eps}}\hspace*{0.4in}
%  \centerline{\epsfxsize=5cm \epsfbox{strainx.eps}}\hspace*{0.4in}
\begin{picture}(3.30,2.6)
%\put(0.10,5.25){\epsfxsize=3.0in \epsfysize=1.78in \epsffile{fig4a.eps} }
%\put(0.10,2.75){\epsfxsize=3.0in \epsfbox{fig4a.eps} }
%\put(0.10,0.25){\epsfxsize=3.0in \epsffile{fig4b.eps} }
%\put(0.10,2.75){\epsfxsize=3.0in \epsfysize=1.78in \epsffile{fig4a.eps} }
  \put(0.10,0.25){\epsfxsize=3.0in \epsfysize=1.78in \epsffile{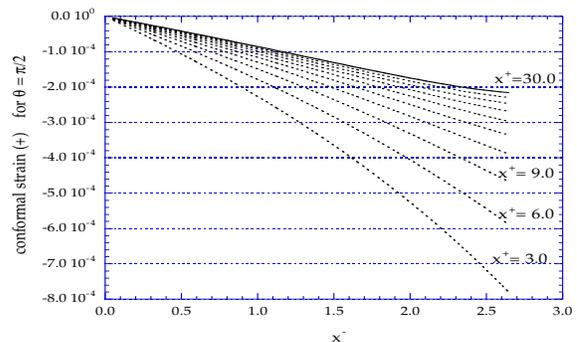} }
\end{picture}
\caption{
Conformal strain $\varepsilon_+$
for $a/m=0.1$ and $r_0=4.0$.
The plot is for the equatorial plane $\theta=\pi/2$,
showing the convergence
of these lines in the $x^+$ direction.
Lines are of $x^+=3.0, 6.0, 9.0, \cdots,$
and $30.0$. We remark that these
lines are not wave-like.}
\label{fig_strain}
\end{figure}

\begin{figure}[t]
\setlength{\unitlength}{1in}
%  \centerline{\epsfxsize=5cm \epsfbox{strain+.eps}}\hspace*{0.4in}
%  \centerline{\epsfxsize=5cm \epsfbox{strainx.eps}}\hspace*{0.4in}
\begin{picture}(3.30,5.2)
\put(0.10,2.75){\epsfxsize=3.0in \epsfysize=1.78in \epsffile{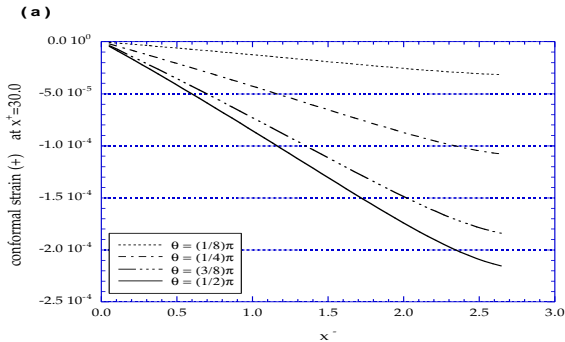} }
\put(0.10,0.25){\epsfxsize=3.0in \epsfysize=1.78in \epsffile{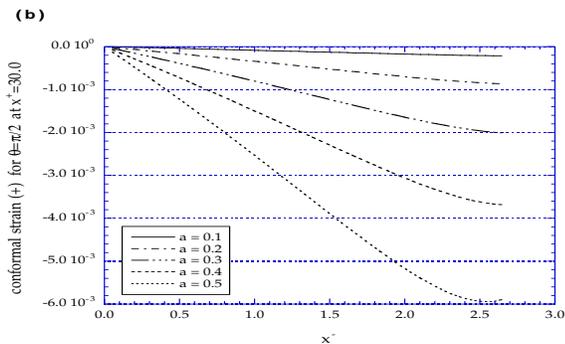} }
\end{picture}
\caption{
Conformal strain $\varepsilon_+$ for the same parameters as
Fig.\ref{fig_strain}.
(a) is $\varepsilon_+$ at $x^+=30.0$ for different $\theta$.
We see that the maximal strain occurs in the equatorial plane,
$\theta=\pi/2$, as expected.
(b) shows the dependence of $\varepsilon_+$ on $a/m$.
The lines are for the data at $x^+=30.0$ for $\theta=\pi
/ 2$. Both solid lines are equivalent with the solid line in
Fig.\ref{fig_strain} (a).}
\label{fig_strain2}
\end{figure}

For the same set of parameters, we also plot the evolution behaviour of
the conformal strain
in Fig.\ref{fig_strain} and Fig.\ref{fig_strain2}.
The cross component, $\varepsilon_\times = \varepsilon_{\theta\phi}$,
is zero in this model,
so only the plus component
$\varepsilon_+ = \varepsilon_{\theta\theta}$ is needed.
The conformal strain
is calculated from (\ref{conformalstrain})
as a function of $x^-$ at constant $x^+$ by setting $\varepsilon_+=0$
at $x^-=0$.
We again observe that $\varepsilon_+$
converges to particular lines (the solid lines) as $x^+$ increases,
again reflecting the conformal variables.
The line of $x^+=30.0$ in Fig.\ref{fig_strain}, therefore,
is close to the waveform for observers
infinitely far from the source.

%%%%%%%%%%%%%%%%%%%%%%%%%%%%%%%%%% figure %%%%%%%%%%%%%%%%
%\begin{figure}[tbh]
%%%%%%%%%%%%%%%%%%%%%%%%%%%%%%%%%% figure %%%%%%%%%%%%%%%%

Note that the horizontal axis in Fig.\ref{fig_strain} is $x^-$ coordinate,
and this is related to $\Delta x^-\approx\sqrt{2}\Delta t$
at large distance, cf.\ (\ref{xpm_from_r}).
Then the unit length $\Delta x^-$ for observers at large distance is about
5 $(m / M_\odot) \mu$ sec,
translated from our units $c=G=1$.
Our plot, therefore, covers a quite short time period
compared with the typical millisecond timescale
of gravitational waves from a Kerr black hole.\footnote
{According to the quasi-normal mode analysis
of the Kerr black hole \cite{Leaver},
the dominant frequencies (fundamental mode corresponding to $l=2$)
of quasi-normal
mode for a 10 $M_\odot$ black hole is between 1.2 kHz (for $a=0$)
and 1.8 kHz (for close to maximally rotating).}
To obtain longer timescales we would have to integrate closer to the horizon,
which causes numerical difficulties due to the infinite redshift.
However, for a dynamically evolving black hole,
the event horizon has finite redshift
and so could lie in the numerical integration region,
allowing evolution to late times.

The magnitude of the conformal strain in Fig.\ref{fig_strain}
is rescaled to the observable strain by (\ref{scale_of_strain}).
We can compare with an example of expected strain
$\epsilon\sim10^{-20}$\cite{closelimit2} for $R=100$ Mpc
by converting our units:
$\epsilon = 3 \times 10^{-27} \varepsilon / (R/100$ Mpc$)$.
This is small enough to validate the quasi-spherical approximation.
The converged conformal strain ($x^+=30$ lines in Fig.\ref{fig_strain})
is increasing as the ingoing coordinate $x^-$ approaches the
black hole horizon.
However, if we extrapolate this magnitude to the horizon
(which will be reached around $x^- \sim 6.0$ for this choice
of parameter),
it is still many orders of magnitude less than expected values.

Moreover, the strain does not behave like a wave
in this case.  This is good news for
future applications of the quasi-spherical approximation, because
the produced spurious waveform is quite different from a
normal gravitational wave.
We also show $\theta$ and $a$ dependencies of $\varepsilon_+$ in
Fig.\ref{fig_strain2}.

%%%%%%%%%%%%%%%%%%%%%%%%%%%%%%%%%% figure %%%%%%%%%%%%%%%%
%\begin{figure}[tbh]
%%%%%%%%%%%%%%%%%%%%%%%%%%%%%%%%%% figure %%%%%%%%%%%%%%%%

\begin{figure}[tbh]
   \unitlength=1in
%       \psfragscanon
%        \psfrag{KerrA}{$a$}
%        \psfrag{r0}{$r_0$}
%        \psfrag{LogEm}{$\log (E(x^+=20)/m) $}
%            \makebox[\columnwidth]{
%             \includegraphics[height=1.4in,keepaspectratio,clip]{fig5_3D.eps}
%         \hspace*{0.4in}
%               }
   \centerline{\epsfxsize=6.5cm \epsfbox{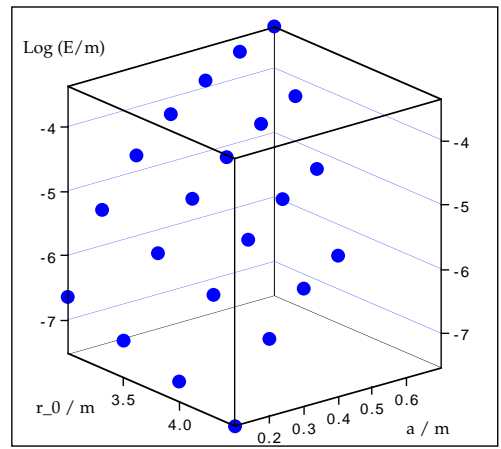}}
   \vspace*{0.2in}
\caption{Logarithmic plot of specific energy $E/m$
due to spurious radiation, as a function of $a/m$ and $r_0/m$.
Energy is measured at $x^+=30$, and the plotted range is
$r_0/m\in[3.0, 4.5]$ and $a/m\in[0.1, 0.7]$.}
\label{fig_result}
\end{figure}

Our final, most conservative check of the quasi-spherical approximation
is to compare the specific energy $E/m$
with the expected specific energy of gravitational waves
from an inspiralling black-hole binary.
The Kerr black-hole space-time seems to be a good example for comparing with
the result of the close-limit approach \cite{closelimit2}.
In Fig.\ref{fig_result}, we plotted the specific energy $E/m$
due to spurious radiation, as a function of $a/m$ and $r_0/m$.
We applied the same grid points and other parameters in numerics with
previous figures, and evaluated $E/m$ at $x^+=30$.
For higher $a$ and larger $r_0$ cases, we could not fill plots in
Fig.\ref{fig_result}.
This is because we kept the resolutions and the
same tolerance for the consistency convergence criteria for all
cases, and these criteria failed for higher $a$  and larger $r_0$.
If we increase the resolutions and/or adjust the
convergence criteria, then we can fill in these missing points also.

Consequently, we observe that
the specific energy $E/m$ increases with $a/m$ and decreases with $r_0/m$,
as expected.
If we compare the amplitude of $E/m$ with Fig.1 of
Khanna {\it et al.} \cite{closelimit2}, then we find that
our values are at least an order
of magnitude smaller than the results
of the close-limit approximation,
up to the range where different versions of the latter diverge.
The fact that
the spurious radiation produced in the quasi-spherical approximation
is quite small
indicates the robustness of this approximation to the general situations.

%%%%%%%%%%%%%%%%%%%%%%%%%%%%%%%%%% figure %%%%%%%%%%%%%%%%
%%%%%%%%%%%%%%%%%%%%%%%%%%%%%%%%%% figure %%%%%%%%%%%%%%%%

%%%%%%%%%%%%%%%%%%%%%%%%%%%%%%%%%%%%%%%%%%%%%%%%%%%%%%%%%%%%%%%%%%%%%%%
\section{Concluding Remarks} \label{sec_conclude}

We tested the quasi-spherical approximation by applying it to
Kerr black holes.  We numerically calculated the strain and energy flux
of the spurious
gravitational radiation produced from this approximation, and showed
that (a) it converges quickly due to our conformal variables, (b)
it does not behave like wave-like oscillations, and (c) the total radiated
energy is at least an
order of magnitude less than the gravitational radiation emission
estimated from coalescing binary black holes, according to the
close-limit approximation\cite{closelimit2}.
We remark that the close-limit approximation
is the only current result which predicts the total amount of
radiation from inspiralling binary black holes.
Numerical results for head-on collisions with appreciable relative momentum
also give similar estimates\cite{baker}.

These results suggest that the spurious radiation does not fatally affect
the gravitational waveform estimation.
It might not affect the waveform estimation at all, and we might
extract its effect to the total energy by subtracting the amount we showed
in Fig.\ref{fig_result}.   These facts directly encourage the robustness of
the quasi-spherical approximation.
Therefore
we are interested in applying this scheme to more general situations,
and/or implementing it as an output routine for full numerical
simulation codes of binary black holes or compact stars,
such as those using the standard 3+1 decomposition of spacetime.
These efforts will be reported elsewhere.

%%%%%%%%%%%%%%%%%%%%%%%%%%%%%%%%%%%%%%%%%%%%%%%%%%%%%%%%%%%%%%%%%%%%%%%
\section*{Acknowledgments}

We thank communications with Abhay Ashtekar, Pablo Laguna,
Luis Lehner, Shinji Mukohyama,
Jorge Pullin and John Stewart.
We appreciate the hospitality of the CGPG group.
%Numerical computations were performed using machines at CGPG.
HS was supported by the Japan Society for the Promotion of Science
as a research fellow abroad.
SAH was supported by the National Science Foundation under award PHY-9800973.

%%%%%%%%%%%%%%%%%%%%%%%%%%%%%%%%%%%%%%%%%%%%%%%%%%%%%%%%%%%%%%%%%%%%%%%

\end{document}